\documentclass[twocolumn,showpacs,preprintnumbers,amsmath,amssymb]{revtex4}
\usepackage[dvips]{graphicx}
\usepackage{amssymb,amsfonts,amsmath}

\begin{document}

\title{The microscopic meaning of grand potential: cluster properties of the one-dimensional lattice gas}
\author{Agata Fronczak}
\email{agatka@if.pw.edu.pl}
\affiliation{Faculty of Physics, Warsaw University of Technology,Koszykowa 75, PL-00-662 Warsaw, Poland}

\date{\today}

\begin{abstract}
We demonstrate, with a concrete example, how the combinatorial approach to a general system of particles, which was introduced in detail in the earlier paper \cite{Fronczak2012}, works and where it enters to provide a genuine extension of results obtainable by more traditional methods of statistical mechanics. To this end, an effort is made to study cluster properties of the one-dimensional lattice gas with nearest neighbor interactions. Three cases: the infinite temperature limit, the range of finite temperatures, and the zero temperature limit are discussed separately, yielding some new results and providing alternative proofs of known results. In particular, the closed-form expression for the grand partition function in the zero temperature limit is obtained, which results in the non-analytic behavior of the grand potential, in accordance with the Yang-Lee theory.
\end{abstract}

\maketitle

\section{Introduction}\label{Sec1}

In our recent paper \cite{Fronczak2012} (hereafter referred to as the paper~I), we have started a new line of theoretical research on imperfect gases and interacting fluids. We have exploited concepts of enumerative combinatorics to deal with equilibrium systems in the infinite volume limit. The approach seems to provide precise mathematical techniques relevant to study phase transitions. In particular, we have shown that the so-called \emph{perfect gas of clusters model} underlying various cluster/droplet theories of phase transitions (see eg. \cite{Sator2003} and references therein) naturally emerges from our approach. In the model, the basic idea is that an imperfect fluid, which is made up of interacting particles, can be considered as an ideal gas of clusters at thermodynamic and chemical equilibrium. There is no potential energy of interaction between clusters and the clusters do not compete with each other for volume. The main conclusion drawn from the earlier paper was that the grand potential (the Landau free energy) of such a clustered system of interacting particles may be considered as the exponential generating function for the number of internal states (thermodynamic probability) of these clusters.

In this paper we will use the combinatorial approach described in the paper I to analyze properties of the one dimensional lattice-gas with nearest neighbor interactions. The aim of this paper is twofold. First, we want to show on a specific case, how our approach to interacting fluids works, and where it enters to provide a genuine extension of results obtainable by more traditional methods. Second, we want to demonstrate with a first concrete example that the approach may provide important insights into microscopic mechanisms which lead to the occurrence of phase transitions. Although phase transitions in the traditional sense are proved not to exist in the one-dimensional lattice gas considered in this paper, its behavior at $T=0K$ seems to have some direct bearing to the problem. To show this, we will focus on the relation between thermodynamics and cluster properties of the lattice gas. The general question of such a relation has been raised by various theories of phase transitions, in which geometric interpretation of the liquid-gas transition was shown to consist in the sudden formation of the macroscopic cluster. In the following, with the example of the one-dimensional lattice gas, we will show that our approach is well suited to handle such problems.

The outline of the paper is as follows. In section~\ref{Sec2} we briefly review results of the paper I.
Section~\ref{Sec3} is devoted to a detailed description of the one-dimensional lattice gas model and its equivalence with the chain of Ising spins in the external magnetic field. In section ~\ref{Sec4}, methods described in Sec.~\ref{Sec2} are used to study cluster properties of the lattice gas. The paper is finalized in section~\ref{Sec5} with some concluding remarks

\section{Microscopic meaning of grand potential}\label{Sec2}

\subsection{Combinatorial approach to a general system of particles}\label{Sec2a}

In the paper~I, we have considered a general system of interacting particles. The thermodynamic state of the system was given by the temperature, $T$, and the chemical potential per molecule, $\mu$. Assuming that the classical treatment is adequate, we have used the grand canonical ensemble to describe the open system in the infinite volume limit, $V\rightarrow\infty$. We have shown that the grand partition function,
\begin{eqnarray}\label{Xi0a}
\Xi(\beta,z)\!&\!=\!&\!\sum_{N=0}^\infty\!z^N\!Z(\beta,N)=1+\sum_{N=1}^{\infty}\!z^N\!Z(\beta,N)\\ \label{Xi0b}\!&\!=\!&\!1+\sum_{N=1}^{\infty}\!z^N\!\int_{0}^\infty\! e^{-\beta E}g(E,N)dE,
\end{eqnarray}
where $\beta=(k_BT)^{-1}$, $z=e^{\beta\mu}$, $Z(\beta,N)$ is the canonical partition function of the considered system with $N$ particles, $Z(\beta,0)=1$ represents the so-called vacuum state, and $g(E,N)$ stands for the density of states, can be written as
\begin{eqnarray}\label{Xi1a}
\Xi(\beta,z)&=&\exp[\;-\beta\;\Phi(\beta,z)]\\ \label{Xi1b} &=&\exp\!\!\left[-\beta\sum_{m=1}^\infty\frac{z^m}{m!}\phi_m(\beta) \right]\\ \label{Xi1c} &=&1+\sum_{N=1}^\infty \frac{z^N}{N!}\;B_{N}(\{w_n(\beta)\})\\ \label{Xi1d} &=&1+\sum_{N=1}^\infty \frac{z^N}{N!}\sum_{k=1}^N B_{N,k}(\{w_n(\beta)\}),
\end{eqnarray}
where the coefficients $w_n(\beta)$ are given by the partial derivatives of the grand thermodynamic potential, $\Phi(\beta,z)$, i.e.
\begin{equation}\label{wn}
w_n(\beta)=-\beta\phi_n(\beta)=-\beta\left.\frac{\partial^n\Phi}{\partial z^n}\right|_{z=0},
\end{equation}
while $B_{N,k}(\{w_n\})=B_{N,k}(w_1,w_2,\dots,w_{N-k+1})$, and $B_N(\{w_n\})=\sum_{k=1}^{N}B_{N,k}(\{w_n\})$ (with $B_0(\{w_n\})\!:=\!0$) represent the so-called incomplete and compete Bell polynomials, respectively.

The partial Bell polynomials in Eq.~(\ref{Xi1d}) are defined as
\begin{equation}\label{Bell}
B_{N,k}(\{w_n(\beta)\})=N!\sum\prod_{n=1}^{N-k+1}\frac{1}{c_n!}\left(\frac{w_n(\beta)}{n!}\right)^{c_n},
\end{equation}
where the summation takes place over all non-negative integers $c_n\geq 0$, such that the constraints $\sum_{n}c_n=k$ and $\sum_{n}nc_n=N$ hold. Given combinatorial meaning of Bell polynomials, we have shown in the paper~I that in the case when the coefficients $w_n(\beta)$ are all non-negative, i.e.
\begin{equation}\label{wng0}
\forall_{n\geq 1}\;w_n(\beta)\geq 0,
\end{equation}
the equality of the two series, Eqs.~(\ref{Xi0b}) and (\ref{Xi1d}), give rise to the following formula:
\begin{equation}\label{main}
\int_0^{\infty}f(k,E)g(E,N)e^{-\beta E}dE=\frac{1}{N!}B_{N,k}(\{w_n(\beta)\}),
\end{equation}
where $f(k,E)$ is the probability that the system having energy $E$ consists of $k$ independent clusters. Also, we have shown that Eq.~(\ref{main}) describes probability that a general system of $N$ interacting particles at temperature $T$, regardless of its energy, consist of $k$ clusters. It has been argued that the formula provides a novel and strictly microscopic understanding of the grand thermodynamic potential, as the exponential generating function for the numbers of internal states, $w_n(\beta)$, of $n$-clusters, i.e.
\begin{eqnarray}\label{Phi}
\Phi(\beta,z)&=&-\frac{1}{\beta}\ln\Xi(\beta,z)\\\label{Phi1}&=&-\frac{1}{\beta}\sum_{n=1}^\infty\frac{w_{n}(\beta)}{n!}z^n.
\end{eqnarray}

Finally, in the earlier paper, we have noted that comparison of Eqs.~(\ref{Xi0a}) and (\ref{Xi1c}) leads to a neat expression for the canonical partition function:
\begin{eqnarray}\label{ZBell1}
Z(\beta,N)&=&\frac{1}{N!}B_N(\{w_n(\beta)\})\\ \label{ZBell2} &=&\frac{1}{N!}\sum_{k=1}^NB_{N,k}(\{w_n(\beta)\}).
\end{eqnarray}
The expression is valid regardless of whether the conditions for the perfect gas of clusters model, Eq.~(\ref{wng0}), are satisfied.

\subsection{Perfect gas of clusters model}\label{Sec2b}

The microscopic meaning of the grand potential described in the paper~I directly relates to the perfect gas of clusters model (which is known from classical theory of simple fluids \cite{Sator2003}). In the model, one deals with a collection of non-interacting clusters, without knowing how one can build them as disjoint sets of interacting particles. In the infinite volume limit, $V\rightarrow\infty$, the pressure, the density, and the cluster size distribution (that is the mean number of clusters of size $n$) are respectively given by:
\begin{eqnarray}
\label{defP}P&=&\frac{1}{\beta V}\sum_{n=1}^\infty Z_n(\beta,V)z^n,\\
\label{defRho}\rho&=&\frac{1}{V}\sum_{n=1}^\infty n Z_n(\beta,V)z^n,
\end{eqnarray}
and
\begin{equation}
\label{defNn}N_n(\beta,V)=Z_n(\beta,V)z^n,
\end{equation}
where $Z_n(\beta,V)$ stands for the partition function which characterizes clusters of size $n$.

The mentioned relation between our results and the classical gas of clusters model \footnote{To justify the name of the model (i.e. perfect gas of clusters), one just needs to write Eq.~(\ref{defP}) as a function of the cluster size distribution, i.e. with the help of Eq.~(\ref{defNn}), i.e. $\beta PV=\sum_{n=1}^\infty N_n(\beta,V)$. The last formula clearly shows that, in the limit of infinite volume, the imperfect gas/fluid can be seen as composed of $\sum_{n=1}^\infty N_n(\beta,V)$ noninteracting and independent clusters.} can be seen by first rewriting Eq.~(\ref{defP}) in the following equivalent form:
\begin{equation}\label{defP1}
\Phi(\beta,z)=-PV=-\frac{1}{\beta}\sum_{n=1}^\infty Z_n(\beta,V)z^n,
\end{equation}
and then comparing Eq.~(\ref{defP1}) with the expression for the grand thermodynamic potential, Eq.~(\ref{Phi1}), which raises the microscopic meaning of $\Phi(\beta,z)$. From the comparison it becomes evident that the conditions given by Eq.~(\ref{wng0}) make our approach to a general system of interacting particles equivalent to the perfect gas of clusters model with
\begin{equation}
\lim_{V\rightarrow\infty} Z_n(\beta,V)=\frac{w_n(\beta)}{n!}.
\end{equation}

\section{One-dimensional lattice gas: Survey of known results}\label{Sec3}

Let us consider a one-dimensional periodic lattice that consists of $V$ sites (in the model, $V$ represents volume) and a collection of $N$ particles. The particles occupy sites of the lattice with the restriction that not more than one particle can occupy a given lattice site and only particles on nearest neighbor sites interact. If we introduce variables $\sigma_i$ for each lattice site $i$, such that $\sigma_i=+1$ if the site is occupied and $\sigma_i=0$ otherwise, then the total energy for a given configuration of particles, $\{\sigma_i\}$, is
\begin{equation}\label{EG}
E_G(\{\sigma_i\})=-\varepsilon \sum_{i=1}^{V}\sigma_i\sigma_{i+1},
\end{equation}
where $-\varepsilon$ is the interaction energy between two neighboring particles, the periodicity of the lattice is imposed by assuming that $\sigma_{V+1}=\sigma_1$, and
\begin{equation}\label{NG}
\sum_{i=1}^{V}\sigma_i=N.
\end{equation}

In order to apply the general method described in Sec.~\ref{Sec2} to investigate the one-dimensional lattice gas, the first task is to find its grand partition function in the infinite volume limit, $V\rightarrow\infty$, i.e.
\begin{eqnarray}\label{XiG1}
\Xi_G(\beta,V,z)&\!=\!&1+\sum_{N=1}^Vz^NZ_G(\beta,V,N)\\\label{XiG2}
&\!=\!& 1+\sum_{N=1}^Vz^N\sum_{\{\sigma_i\}^*}e^{-\beta E_G(\{\sigma_i\})}\\\label{XiG3} &\!=\!& \sum_{\{\sigma_i\}} \exp\left[\beta\varepsilon\sum_{i=1}^V\sigma_i\sigma_{i+1}+\beta\mu\sum_{i=1}^V\sigma_i\right],\;\;\;\;\;
\end{eqnarray}
where $Z_G(\beta,V,N)$ in Eq.~(\ref{XiG1}) is the canonical partition function for this gas and  the starred configurations, $\{\sigma_i\}^*$, in the second sum of~Eq.~(\ref{XiG2}) are those for which the condition given by Eq.~(\ref{NG}) holds. To carry on we exploit the mathematical equivalence between the grand partition function for the lattice gas and the canonical partition function for the Ising model in the external magnetic field \cite{1952LeePhysRev}.

The one dimensional Ising model consists of a chain of $V$ spins, $s_i=\pm 1$, with the interaction energy in a given configuration, $\{s_i\}$, given by
\begin{equation}\label{EI}
E_I(\{s_i\})=-J \sum_{i=1}^{V}s_is_{i+1}-H\sum_{i=1}^Vs_i,
\end{equation}
where $J$ is the coupling constant between nearest neighbors, $H$ is the external magnetic field, and $s_{V+1}=s_1$. The canonical partition function for the model can be written as
\begin{eqnarray}\label{ZI1}
Z_I(\beta,V,H)&\!=\!&\sum_{\{s_i\}}e^{-\beta E_I(\{s_i\})}\\\label{ZI2}&\!=\!& \sum_{\{s_i\}}\exp\!\left[\!\beta J\sum_{i=1}^Vs_is_{i+1}+\beta H\sum_{i=1}^Vs_i\!\right].\;\;\;\;\;
\end{eqnarray}

To show that Eqs.~(\ref{XiG3}) and~(\ref{ZI2}) are in fact equivalent one has merely to note that the variables $\sigma_i$ can be obtained from the variables $s_i$ by writing
\begin{equation}\label{s0}
\sigma_i=\frac{s_i+1}{2}.
\end{equation}
Substituting (\ref{s0}) into (\ref{XiG3}) one gets
\begin{equation}\label{XiG4}
\Xi_G(\beta,V,z)=e^{\beta C_GV}Z_I(\beta,V,H_G),
\end{equation}
where
\begin{equation}\label{s1}
C_G=\frac{\varepsilon}{4}+\frac{\mu}{2},
\end{equation}
and
\begin{equation}\label{XiG5}
Z_I(\beta,V,H_G)\!=\!\sum_{\{s_i\}}\!\exp\left[\beta J_G\sum_{i=1}^Vs_is_{i+1}+\beta H_G\sum_{i=1}^Vs_i\right]
\end{equation}
is the Ising partition function, Eq.~(\ref{ZI2}), with
\begin{equation}\label{s2}
J_G=\frac{\varepsilon}{4},\;\;\;\;\;\;H_G=\frac{\varepsilon+\mu}{2}.
\end{equation}

Now, putting into Eq.~(\ref{XiG4}) the well-known exact formula for the partition function of the closed Ising chain of $V$ spins in the external magnetic field \cite{bookBaxter,bookStauffer}, i.e.
\begin{eqnarray}\label{XiG6a}
Z_I(\beta,\!V,\!H_G)&\!=\!&\lambda_{+}(\beta,\!H_G)^V+\lambda_{-}(\beta,\!H_G)^V\\ \label{XiG6b} \!&\stackrel{V\rightarrow\infty}{\simeq}\!&\lambda_{+}(\beta,\!H_G)^V,
\end{eqnarray}
where
\begin{equation}\label{ZI4}
\lambda_\pm\!(\beta,\!H_G)\!=\!e^{\beta J_G}\!\!\left(\!\!\cosh(\beta H_G)\!\pm\!\sqrt{\sinh^2(\beta H_G)\!+\!e^{-4\beta J_G}}\!\!\right)\!\!,
\end{equation}
the grand partition function for the one-dimensional lattice gas with nearest-neighbor interactions becomes, for $V\rightarrow\infty$,
\begin{equation}\label{XiG7}
\Xi_G(x,V,z)=\left(x\sqrt{z}\;\lambda_{+}\!(x,z)\right)^V,
\end{equation}
where
\begin{equation}\label{defxz1}
x=e^{\beta\varepsilon/4},\;\;\;\;\;z=e^{\beta\mu},
\end{equation}
and
\begin{equation}\label{defxz2}
\lambda_{\pm}\!(x,\!z)\!=\!\frac{x^3\sqrt{z}}{2}+\!\frac{1}{2x\sqrt{z}}\pm \sqrt{\frac{x^6z}{4}+\!\frac{1}{4x^2z}+\!\frac{1}{x^2}-\!\frac{x^2}{2}}
\end{equation}
corresponds to Eq.~(\ref{ZI4}) but written in the new variables $x$ and $z$. Finally, inserting Eq.~(\ref{XiG7}) into Eq.~(\ref{Phi}) one gets free energy of the one-dimensional lattice gas with nearest neighbor interactions in the infinite volume limit,
\begin{equation}\label{PhiG}
\Phi_G(x,V,z)=-\frac{V}{\beta}\ln\left(x\sqrt{z}\;\lambda_{+}\!(x,z)\right).
\end{equation}
Further in this paper, in Sec.~\ref{Sec4}, results of this section will be used to analyze cluster properties of the lattice gas model.


\section{Clusters in lattice-gas model}\label{Sec4}

In this section, the combinatorial approach described in Sec.~\ref{Sec2} will be used to study properties of the one-dimensional lattice-gas model. In what follows, three cases: i. the infinite temperature, ii. the range of finite temperatures, and iii. the zero temperature limit, will be discussed separately.

\subsection{Infinite temperature limit}\label{Sec4a}

In the infinite temperature limit one has
\begin{equation}\label{betaT8K}
\lim_{T\rightarrow\infty}\beta=0.
\end{equation}
Therefore, since $\varepsilon=\mbox{const}$ one gets, see Eq.~(\ref{defxz1}),
\begin{equation}\label{xT8K}
\lim_{T\rightarrow\infty}x=1,
\end{equation}
and the expression for the grand partition function, Eq.~(\ref{XiG7}), simplifies to
\begin{equation}\label{XiT8K}
\lim_{T\rightarrow\infty}\Xi_G(x,V,z)=(1+z)^V.
\end{equation}
Accordingly, the grand potential becomes
\begin{equation}\label{PhiT8K}
\lim_{T\rightarrow\infty}\Phi_G(x,V,z)=-\frac{V}{\beta}\ln(1+z),
\end{equation}

Successive derivatives, $\phi_n$, of the grand potential with respect to $z$ and evaluated at $z=0$, Eq.~(\ref{wn}), give the following closed-form expression for the Bell polynomials' coefficients:
\begin{equation}\label{wnT8K}
\lim_{T\rightarrow\infty}w_n(\beta)=V(n-1)!(-1)^{n-1}.
\end{equation}
From the last expression, it is obvious that the parameters do not satisfy the conditions under which the gas can be considered as the ideal gas of clusters, cf.~Eq.~(\ref{wng0}). By definition, lattice-gas clusters are sets of neighboring particles. There is no energy of interaction between such clusters, cf.~Eq.~(\ref{EG}). However, due to the so-called excluded volume effect, lattice-gas clusters in some sense interact with each other, even if there is no direct interaction between their particles. The effect has its origin in the cluster counting problem, and consists in the fact that two clusters of a given size once defined cannot approach close enough to one another to be counted, under the definition, as a single cluster. Consequently, only at small densities, the lattice-gas can be considered as a perfect gas of clusters.

Nevertheless, the case of the one-dimensional lattice gas in the infinite temperature limit allows to perform a direct validation of our combinatorial approach. By inserting the coefficients $\{w_n(\beta)\}$, Eq.~(\ref{wnT8K}), into the expression for the canonical partition function, Eq.~(\ref{ZBell2}), after some algebra one gets
\begin{eqnarray}\label{ZT8Ka}
\lim_{T\rightarrow\infty}Z_G(\beta,V,N) \!&\!=\!&\!\sum_{k=1}^N \frac{V^k(\!-1\!)^{N-k}}{N!}B_{N,k}(\{(\!n\!-\!1\!)!\})\;\;\;\;\;\; \\ \label{ZT8Kb} \!&\!=\!&\! \sum_{k=1}^N\!\frac{V^k(\!-1\!)^{N-k}}{N!}|s(N,k)|\!=\!\!{V \choose N},
\end{eqnarray}
where $|s(N,k)|$ represents the signless (or unsigned) Stirling number of the first kind,  ${V \choose N}$ stands for the binomial coefficient, and where some basic combinatorial identities have been used, including: i. properties of Bell polynomials (\cite{bookComtet}, pp.~133--137), i.e.
\begin{equation}\label{Bell1}
B_{N,k}(\{ab^nx_n\})=a^kb^NB_{N,k}(\{x_n\}),
\end{equation}
and
\begin{equation}\label{Bell2}
B_{N,k}(\{(n-1)!\})=|s(N,k)|,
\end{equation}
and ii. identities involving Stirling numbers of the first kind \cite{WolframStirling}, i.e. the relation between signed and unsigned Stirling numbers,
\begin{equation}\label{Stirling1}
|s(N,k)|=(-1)^{N-k}s(N,k),
\end{equation}
and its generating function,
\begin{equation}\label{Stirling2}
{V\choose N}=\sum_{k=1}^N\frac{V^k}{N!}s(N,k).
\end{equation}

The obtained result, Eq.~(\ref{ZT8Kb}),
\begin{equation}\label{ZT8Kc}
\lim_{T\rightarrow\infty}Z_G(\beta,V,N) ={V\choose N},
\end{equation}
is exactly the expected one. In the infinite temperature limit, all accessible microstates are equally probable. Therefore, the canonical partition function is just the number of microstates allowed, i.e. the number of ways to choose positions for $N$ particles from the available $V$ positions.

\subsection{Finite temperatures}\label{SecT8K}

In the range of finite temperatures, i.e. for
\begin{equation}\label{betaT}
0<\beta<\infty,\;\;\;\;\;\;\;1<x<\infty,
\end{equation}
and for small particle densities,
\begin{equation}\label{rhoT}
\lim_{V\rightarrow\infty}\frac{N}{V}\ll 1,
\end{equation}
one can directly test Eq.~(\ref{main}), which describes the probability that the gas of $N$ particles, regardless of its energy, consists of $k$ clusters. The expression mentioned holds exactly only for the ideal gas of clusters. In this case, however, although in general the lattice-gas does not satisfy the conditions specified by Eq.~(\ref{wng0}), but given small particle densities the coefficients $w_n(\beta)$ are nonnegative for a reasonable range of cluster sizes, $n$, and Eq.~(\ref{main}) seems to provide good approximation for cluster statistics.

\begin{figure}
\includegraphics[width=7.5cm]{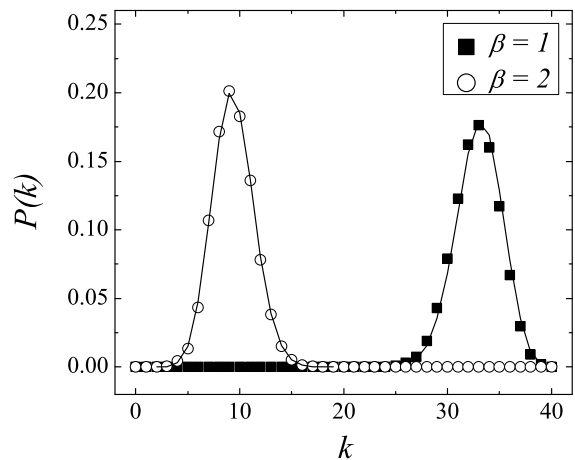}
\caption{\label{fig1}Comparison of probability distributions for the number of clusters, $k$, in the one-dimensional lattice gas obtained from Monte Carlo simulations of $N=40$ particles on $V=8000$ sites (scattered points) and theoretical distributions $P(k)$ given by Eq.~(\ref{Pk}) (solid lines). Numerical simulations were done using Metropolis algorithm with the coupling energy between two neighboring particles $\varepsilon=4$, cf.~Eq.~(\ref{EG}), and for two different values of the inverse temperature, $\beta=1$ and $2$.}
\end{figure}

The coefficients $w_n(\beta)$ of Bell polynomials in Eq.~(\ref{main}),
\begin{eqnarray}\label{wnT1}
w_1(\beta)\!&=&\!V,\\\label{wnT2}
w_2(\beta)\!&=&\!V(2x^4-3),\\\label{wnT3}
w_3(\beta)\!&=&\!V(6x^8-12x^4+10) \\ \nonumber \dots \\\label{wnTn} w_n(\beta)\!&=&\!V\left(n!x^{4(n-1)}-\dots\right),
\end{eqnarray}
and also the polynomials themselves can be easily calculated using Mathematica (Wolfram, Inc.). The normalized probability distribution, $P(k)$, for the number of clusters, $k$, i.e. the right hand side of Eq.~(\ref{main}) divided by Eq.~(\ref{ZBell1}),
\begin{equation}\label{Pk}
P(k)=\frac{B_{N,k}(\{w_n(\beta)\})}{B_{N}(\{w_n(\beta)\})},
\end{equation}
is shown in Fig.~\ref{fig1} together with Monte Carlo simulations of the one-dimensional conserved-order-parameter Ising model \cite{bookNewman}, which is often used to study properties of lattice gases.

In the figure, one can see that for $N,V=\mbox{const}$ the average number of clusters, and consequently also the average cluster size strongly depend on temperature. At lower temperatures (i.e. for higher vales of $\beta$) particles try to stay in clusters. Configurations, $\{\sigma_i\}$, with less number of large clusters are more likely to be observed with respect to configurations with a large number of small clusters which are typical of higher temperatures.

\subsection{Zero temperature limit}\label{SecT0K}

\subsubsection{Non-analyticity of the grand partition function}\label{SecT0Ka}

The van Hove's theorem \cite{bookLieb, bookDomb} states that the limiting free energy, Eq.~(\ref{Phi1}), characterizing one-dimensional gases with short range interactions (such as the one considered in this study), is an analytic function for all real positive values of $T$. This statement excludes phase transitions for $T>0$K. The theorem, however, does not say anything about analyticity of the free energy at $T=0$K. In fact, the grand potential given by Eq.~(\ref{PhiG}) is non-analytic at $T=0$K, i.e. for
\begin{equation}\label{betaT0K}
\lim_{T\rightarrow 0K}\beta=\infty,
\end{equation}
and
\begin{equation}\label{xT0K}
\lim_{T\rightarrow 0K}x=\infty.
\end{equation}
Obviously, this non-analyticity is of the same mathematical nature as the one observed in the free energy of the one-dimensional Ising model at zero temperature, when the magnetic field, $H$, goes to zero.

At low temperatures, $T\rightarrow 0$K, one finds that  Eq.~(\ref{ZI4}) becomes \cite{bookStauffer}
\begin{eqnarray}\nonumber
\lambda_\pm\!(\beta,\!H_G)\!&\!=\!&\!e^{\beta J_G}\!\left(\!\cosh(\beta H_G)\!\pm\!|\sinh(\beta H_G)|(\!1\!+\!O[e^{-4\beta J_G}])\right)\\ &\simeq\!&\!e^{\beta(J_G\pm|H_G|)}\label{ZI5b}.
\end{eqnarray}
Rewriting the last expression in the variables $x$ and $z$, cf. Eq.~(\ref{defxz1}), after some algebra one gets
\begin{equation}\label{ZI6}
\lim_{T\rightarrow 0K}\lambda_\pm(x,z)=xe^{\pm|\ln(x^2\sqrt{z})|}.
\end{equation}
Correspondingly, the resulting grand partition function for $V\rightarrow\infty$, Eq.~(\ref{XiG7}), can be written as follows
\begin{equation}\label{XiGT0K}
\lim_{T\rightarrow 0K}\Xi_G(x,V,z)=\left(e^{\ln(x^2\!\sqrt{z})+|\ln(x^2\!\sqrt{z})|}\right)^V.
\end{equation}
The expression reveals the non-analytic behavior at
\begin{equation}\label{T0K}
\lim_{T\rightarrow 0K}x^4z=1,
\end{equation}
since
\begin{equation}\label{XiGT0Ka}
\lim_{x^4\!z\rightarrow 1^{-}}\Xi_G(x,V,z)=1,
\end{equation}
while
\begin{equation}\label{XiGT0Kb}
\lim_{x^4\!z\rightarrow 1^{+}}\Xi_G(x,V,z)=x^{4V}z^V.
\end{equation}

\subsubsection{Cluster properties}\label{SecT0Kb}

The non-analytic behavior of the grand partition function just described has a very clear and convincing interpretation in terms of our combinatorial approach. Comparing Eqs.~(\ref{XiGT0Ka}) and (\ref{XiGT0Kb}) with Eq.~(\ref{Xi0a}) one gets the following expressions for the canonical partition functions:
\begin{equation}\label{XiGT0Kaa}
\lim_{x^4\!z\rightarrow 1^{-}}Z_G(\beta,V,N)=\delta_{N,0},
\end{equation}
and
\begin{equation}\label{XiGT0Kbb}
\lim_{x^4\!z\rightarrow 1^{+}}Z_G(\beta,V,N)=x^{4N}\delta_{N,V},
\end{equation}
where $\delta_{i,j}$ is the Kronecker delta.

The last two expressions show that in the considered case of the zero temperature limit, the one-dimensional lattice gas with nearest neighbor interactions can be observed in only two particle configurations. The first configuration,
\begin{equation}\label{siT0Ka}
\forall_{1\leq i\leq V}\;\sigma_i= 0.
\end{equation}
is consistent with Eqs.~(\ref{XiGT0Ka}) and (\ref{XiGT0Kaa}), and describes system with no particles (the so-called vacuum state). The second configuration,
\begin{equation}\label{siT0Kb}
\forall_{1\leq i\leq V}\;\sigma_i= 1.
\end{equation}
results from Eqs.~(\ref{XiGT0Kb}) and (\ref{XiGT0Kbb}), and corresponds to the spanning cluster of size $V$.

To proceed with understanding the combinatorial approach described in Sec.~\ref{Sec2}, it is instructive to analyze Eqs.~(\ref{XiGT0Kaa}) and (\ref{XiGT0Kbb}) with the help of the general formula for the canonical partition function, Eq.~(\ref{ZBell2}). Using properties of Bell polynomials, i.e. \begin{equation}\label{Bell4}
B_{N,k}(0,0,\dots,0,x_j,0,\dots)=\delta_{N\!,jk}\;\frac{(jk)!}{k!(j!)^k}x_j^k,
\end{equation}
one can show that the case when $x^4z$ approaches $1$ from the left hand side, Eq.~(\ref{XiGT0Kaa}), corresponds to:
\begin{equation}\label{wnT0Ka}
\forall_{n\geq 1}\;w_n(\beta)= 0,
\end{equation}
while the case when $x^4z$ approaches $1$ from the right hand side, Eq.~(\ref{XiGT0Kbb}), amounts to:
\begin{equation}\label{wnT0Kb1}
\forall_{n\geq 1;n\neq V}\;w_n(\beta)= 0,
\end{equation}
and
\begin{equation}\label{wnT0Kb2}
w_V(\beta)=V!\;x^{4V}.
\end{equation}
From Eqs.~(\ref{wnT0Ka}) - (\ref{wnT0Kb2}) it is evident that the considered limiting behavior meets the conditions specified in Eq.~(\ref{wng0}). It validates our cluster-based combinatorial approach to a general system of interacting particles and also highlights its potential usefulness for the theory of phase transitions.

\subsubsection{Zeros of the grand partition function}

In the zero temperature limit, i.e. for $x\rightarrow\infty$, Eq.~(\ref{wnTn}) can be approximated by
\begin{equation}\label{wnTn0K}
\forall_{n\geq 1}\;w_n(\beta)=Vn!x^{4(n-1)}.
\end{equation}
Inserting the last expression into Eq.~(\ref{ZBell2}) for the canonical partition function one gets:
\begin{eqnarray}\label{ZG1}
Z_G(x,V,N)&=&\frac{1}{N!}\sum_{k=1}^NB_{N,k}(\{Vn!x^{4(n-1)}\})\\ \label{ZG2}&=& \frac{1}{N!}\sum_{k=1}^NV^k(x^{4})^{(N-k)}B_{N,k}(\{n!\})
\\ \label{ZG3}&=& \frac{1}{N!}\sum_{k=1}^NV^k(x^{4})^{(N-k)} L(N,k),
\end{eqnarray}
where Eq.~(\ref{Bell1}) has been used, and where $L(N,k)$ represent the so-called Lah numbers, which are defined as follows
\begin{equation}\label{lah}
L(N,k)={N-1 \choose k-1}\frac{N!}{k!}=B_{N,k}(\{n!\}).
\end{equation}

Lah numbers have an interesting meaning in combinatorics: they count the number of ways a set of $N$ elements can be partitioned into $k$ nonempty linearly ordered subsets. The combinatorial meaning of $L(N,k)$ allows a direct understanding of the canonical partition function, $Z_G(x,V,z)$, as given by Eq.~(\ref{ZG3}). In short, Eq.~(\ref{ZG3}) states that there are $L(N,k)$ particle configurations in which $N$ particles can be arranged into $k$ linear clusters. All such configurations have the same energy, Eq.~(\ref{EG}), and, therefore, also the same Boltzmann factor, $x^{4(N-k)}$. Finally, the factor $V^k$ is due to cluster arrangement along the one-dimensional lattice, in which one assumes that every cluster may start at the same lattice site. Obviously, the arrangement factor does not take into account the excluded-volume effect. For the reason it is only correct in the zero temperature limit.

Now, putting the canonical partition function given by Eq.~(\ref{ZG3}) into the general formula for the grand partition function, Eq.~(\ref{Xi0a}), one gets
\begin{eqnarray}\label{XiG8a}
\Xi_G(x,z)\!&\!=\!&\!1\!+\!\!\sum_{N=1}^\infty\sum_{k=1}^N\frac{\left(zx^4\right)^N }{N!}\! \left(\frac{V}{x^4}\right)^k\!L(N,k)\\\label{XiG8b}\!&\!=\!&\!\exp\left[\frac{Vz}{1-x^4z}\right],
\end{eqnarray}
where the infinite expansion involving Lah numbers has been used (see \cite{bookComtet}, p.~156, or \cite{bookRoman}, pp.~108-113), i.e.
\begin{equation}\label{lah1}
\exp\left[\frac{tu}{1-t}\right]=1+\sum_{N=1}^\infty\sum_{k=1}^N\frac{t^Nu^k}{k!}{N-1 \choose k-1}.
\end{equation}

The grand partition function, $\Xi_G(x,z)$, as calculated above applies to the one-dimensional lattice gas in the thermodynamic limit. In Eq.~(\ref{XiG8b}), volume plays the role of an extensive factor of the grand potential, Eq.~(\ref{Phi}),
\begin{equation}\label{PhiG1}
\Phi_G(x,z)=-\frac{z}{\beta(1-x^4z)}\;V,
\end{equation}
rather than an independent variable. It also must be noted that, in the considered zero temperature limit, $x\rightarrow \infty$, the grand potential per unit volume, which defines pressure, Eq.~(\ref{defP1}),
\begin{equation}\label{defP2}
P=-\frac{z}{\beta(1-x^4z)},
\end{equation}
reveals a non-analytic dependence on $z$ at $x^4z\rightarrow 1$. The non-analyticity translates into the root of the grand partition function, Eq.~(\ref{XiG8b}), for $z=0$.

The significance of the zeros of the grand partition function was first pointed out by Yang and Lee \cite{1952YangPhysRev}, who showed that a phase transition in the sense of a non-analytic dependence of $P$ on $z$ for physical (i.e. real and positive) values of $z$ can only occur when $\Xi_G(x,z)=0$. In the one-dimensional lattice gas with nearest neighbor interactions analyzed in this study, the only root of the grand partition function is for $z=0$, i.e. it does not occur on the positive $z$ axis. Therefore, one claims that phase transitions in the traditional sense do not exist in the considered gas, although its behavior in the zero temperature limit has some direct bearing to the problem of phase transitions.

\subsubsection{Supplementary remark}

It is interesting to see that Eq.~(\ref{PhiG1}) can be derived most easily by inserting Eq.~(\ref{wnTn0K}) into Eq.~(\ref{Phi1}) and assuming that $x^4z<1$:
\begin{equation}\label{PhiG2}
\Phi_G(x,z)= -\frac{V}{\beta x^4}\sum_{n=1}^\infty(x^4z)^n=-\frac{z}{\beta(1-x^4z)}\;V.
\end{equation}
There is a problem with proceeding in this way, however. At variance with our previous results, the obtained expression is not justified for $x^4z>1$. There is no such a problem if one uses our combinatorial approach. The reason is that the approach described in Sec.~\ref{Sec2} is based on the concept of \emph{formal power series} \cite{bookComtet, bookWilf}.

In mathematics, formal power series are a generalization of polynomials as formal objects. A formal power series is an object that just records a sequence of coefficients. One may think of such a series as a power series in which one ignores questions of convergence. However, formal power series still allow one to employ much of the analytical machinery of \emph{normal} power series - especially in settings which do not have natural notions of convergence. We believe that this perspective makes our combinatorial approach to equilibrium statistical mechanics peculiarly well suited to handle problems with convergence, which are often encountered in the theory of phase transitions.

\section{Summary}\label{Sec5}

The aim of this paper was to demonstrate, with a concrete example, how our general combinatorial approach to interacting fluids works. To this end, the one-dimensional lattice gas with nearest neighbor interactions has been considered. Exploiting the mathematical equivalence between the grand partition function for the gas and the canonical partition function for the one-dimensional Ising model in the external magnetic field, cluster properties of the former has been explored. Three cases: the infinite temperature limit, the range of finite temperatures, and the zero temperature limit has been discussed separately. In particular, in the range of finite temperature and for small particle densities the normalized probability distribution for the number of clusters has been found, and the non-analytic behavior of the grand potential in the zero temperature limit, having a direct bearing on phase transitions has been analyzed. The investigation of the zero temperature limit for the gas, has allowed us to remark on the method of formal power series which is behind our approach, and which, we believe, makes the approach peculiarly well suited to handle problems covered by the theory of phase transitions.

In this study, the main purpose was to validate general results of our earlier paper. Apart from this purpose, however, we have made interesting additions to the still-developing theory of one-dimensional lattice gases (see e.g. \cite{T2011,T2005,T2002,T2001}), which has proven useful in studying many natural phenomena in nanophysics, surface science, and biophysics (see e.g. \cite{E2000,E2004,E2003,E1985}).

\acknowledgments
I am indebted to my husband, Dr. Piotr Fronczak, for the help in performing Monte Carlo simulations, results of which are shown in Fig.~\ref{fig1}. I also would like to thank him for his patience with respect to my struggling with relations between physics and combinatorics, which has begun in 2010, just after our second son was born, and lasts incessantly up to this day.

The work has been supported from the internal funds of the Faculty of Physics at Warsaw University of Technology and from the Ministry of Science and Higher Education in Poland (national three-year scholarship for outstanding young scientists 2010-2013).


\begin{thebibliography}{21}
\expandafter\ifx\csname natexlab\endcsname\relax\def\natexlab#1{#1}\fi
\expandafter\ifx\csname bibnamefont\endcsname\relax
  \def\bibnamefont#1{#1}\fi
\expandafter\ifx\csname bibfnamefont\endcsname\relax
  \def\bibfnamefont#1{#1}\fi
\expandafter\ifx\csname citenamefont\endcsname\relax
  \def\citenamefont#1{#1}\fi
\expandafter\ifx\csname url\endcsname\relax
  \def\url#1{\texttt{#1}}\fi
\expandafter\ifx\csname urlprefix\endcsname\relax\def\urlprefix{URL }\fi
\providecommand{\bibinfo}[2]{#2}
\providecommand{\eprint}[2][]{\url{#2}}

\bibitem[{\citenamefont{Fronczak}(2012)}]{Fronczak2012}
\bibinfo{author}{\bibfnamefont{A.}~\bibnamefont{Fronczak}},
  \bibinfo{howpublished}{arXiv:1205.4986v1 [physics.gen-ph]}
  (\bibinfo{year}{2012}), \bibinfo{note}{submitted to Phys. Rev. E}.

\bibitem[{\citenamefont{Sator}(2003)}]{Sator2003}
\bibinfo{author}{\bibfnamefont{N.}~\bibnamefont{Sator}},
  \bibinfo{journal}{Phys. Rep.} \textbf{\bibinfo{volume}{376}},
  \bibinfo{pages}{1} (\bibinfo{year}{2003}).

\bibitem[{\citenamefont{Lee and Yang}(1952)}]{1952LeePhysRev}
\bibinfo{author}{\bibfnamefont{T.~D.} \bibnamefont{Lee}} \bibnamefont{and}
  \bibinfo{author}{\bibfnamefont{C.~N.} \bibnamefont{Yang}},
  \bibinfo{journal}{Phys. Rev.} \textbf{\bibinfo{volume}{87}},
  \bibinfo{pages}{410} (\bibinfo{year}{1952}).

\bibitem[{\citenamefont{Baxter}(1982)}]{bookBaxter}
\bibinfo{author}{\bibfnamefont{R.~J.} \bibnamefont{Baxter}},
  \emph{\bibinfo{title}{Exactly Solved Models in Statitstical Mechanics}}
  (\bibinfo{publisher}{Academic Press}, \bibinfo{address}{London},
  \bibinfo{year}{1982}), chap.~\bibinfo{chapter}{2}.

\bibitem[{\citenamefont{Chowdhury and Stauffer}(2000)}]{bookStauffer}
\bibinfo{author}{\bibfnamefont{D.}~\bibnamefont{Chowdhury}} \bibnamefont{and}
  \bibinfo{author}{\bibfnamefont{D.}~\bibnamefont{Stauffer}},
  \emph{\bibinfo{title}{Principles of Equilibrium Statistical Mechanics}}
  (\bibinfo{publisher}{Willey-VCH}, \bibinfo{address}{Weinheim},
  \bibinfo{year}{2000}), pp. \bibinfo{pages}{311--321}.

\bibitem[{\citenamefont{Comtet}(1974)}]{bookComtet}
\bibinfo{author}{\bibfnamefont{L.}~\bibnamefont{Comtet}},
  \emph{\bibinfo{title}{Advanced Combinatorics: The Art of Finite and Infinite
  Expansions}} (\bibinfo{publisher}{Reidel Publishing Company},
  \bibinfo{address}{Dordrecht}, \bibinfo{year}{1974}).

\bibitem[{\citenamefont{Weisstein}(2012)}]{WolframStirling}
\bibinfo{author}{\bibfnamefont{E.~W.} \bibnamefont{Weisstein}},
  \emph{\bibinfo{title}{Stirling number of the first kind}},
  \bibinfo{howpublished}{From MathWorld--A Wolfram Web Resource}
  (\bibinfo{year}{2012}).

\bibitem[{\citenamefont{Newman and Barkema}(2006)}]{bookNewman}
\bibinfo{author}{\bibfnamefont{M.~E.~J.} \bibnamefont{Newman}}
  \bibnamefont{and} \bibinfo{author}{\bibfnamefont{G.~T.}
  \bibnamefont{Barkema}}, \emph{\bibinfo{title}{Monte Carlo Methods in
  Statistical Physics}} (\bibinfo{publisher}{Oxford University Press},
  \bibinfo{address}{Oxford}, \bibinfo{year}{2006}), chap.~\bibinfo{chapter}{5}.

\bibitem[{\citenamefont{Lieb and Mattis}(1966)}]{bookLieb}
\bibinfo{author}{\bibfnamefont{E.~H.} \bibnamefont{Lieb}} \bibnamefont{and}
  \bibinfo{author}{\bibfnamefont{D.~C.} \bibnamefont{Mattis}},
  \emph{\bibinfo{title}{Mathematical Physics in One Dimension}}
  (\bibinfo{publisher}{Academic Press}, \bibinfo{address}{London},
  \bibinfo{year}{1966}), chap.~\bibinfo{chapter}{1}.

\bibitem[{\citenamefont{Griffiths}(1972)}]{bookDomb}
\bibinfo{author}{\bibfnamefont{R.~B.} \bibnamefont{Griffiths}}, in
  \emph{\bibinfo{booktitle}{Phase Transitions and Critical Phenomena}}, edited
  by \bibinfo{editor}{\bibfnamefont{C.}~\bibnamefont{Domb}} \bibnamefont{and}
  \bibinfo{editor}{\bibfnamefont{M.~S.} \bibnamefont{Green}}
  (\bibinfo{publisher}{Academic Press}, \bibinfo{address}{London},
  \bibinfo{year}{1972}), vol.~\bibinfo{volume}{1}, chap.~\bibinfo{chapter}{2},
  \bibinfo{edition}{1st} ed.

\bibitem[{\citenamefont{Roman}(1984)}]{bookRoman}
\bibinfo{author}{\bibfnamefont{S.}~\bibnamefont{Roman}},
  \emph{\bibinfo{title}{The Umbral Calculus}} (\bibinfo{publisher}{Academic
  Press}, \bibinfo{address}{London}, \bibinfo{year}{1984}), pp.
  \bibinfo{pages}{108--113}.

\bibitem[{\citenamefont{Yang and Lee}(1952)}]{1952YangPhysRev}
\bibinfo{author}{\bibfnamefont{C.~N.} \bibnamefont{Yang}} \bibnamefont{and}
  \bibinfo{author}{\bibfnamefont{T.~D.} \bibnamefont{Lee}},
  \bibinfo{journal}{Phys. Rev.} \textbf{\bibinfo{volume}{87}},
  \bibinfo{pages}{404} (\bibinfo{year}{1952}).

\bibitem[{\citenamefont{Wilf}(1990)}]{bookWilf}
\bibinfo{author}{\bibfnamefont{H.~S.} \bibnamefont{Wilf}},
  \emph{\bibinfo{title}{Generatingfunctionology}} (\bibinfo{publisher}{Academic
  Press}, \bibinfo{address}{New York}, \bibinfo{year}{1990}).

\bibitem[{\citenamefont{Mirabella and Aldao}(2011)}]{T2011}
\bibinfo{author}{\bibfnamefont{D.~A.} \bibnamefont{Mirabella}}
  \bibnamefont{and} \bibinfo{author}{\bibfnamefont{C.~M.} \bibnamefont{Aldao}},
  \bibinfo{journal}{J. Stat. Mech.} \textbf{\bibinfo{volume}{P03011}}
  (\bibinfo{year}{2011}).

\bibitem[{\citenamefont{Yilmaz and Zimmermann}(2005)}]{T2005}
\bibinfo{author}{\bibfnamefont{M.~B.} \bibnamefont{Yilmaz}} \bibnamefont{and}
  \bibinfo{author}{\bibfnamefont{F.~M.} \bibnamefont{Zimmermann}},
  \bibinfo{journal}{Phys. Rev. E} \textbf{\bibinfo{volume}{71}},
  \bibinfo{pages}{026127} (\bibinfo{year}{2005}).

\bibitem[{\citenamefont{Campi et~al.}(2002)\citenamefont{Campi, Krivine, and
  Krivine}}]{T2002}
\bibinfo{author}{\bibfnamefont{X.}~\bibnamefont{Campi}},
  \bibinfo{author}{\bibfnamefont{H.}~\bibnamefont{Krivine}}, \bibnamefont{and}
  \bibinfo{author}{\bibfnamefont{J.}~\bibnamefont{Krivine}},
  \bibinfo{journal}{Physica A} \textbf{\bibinfo{volume}{320}},
  \bibinfo{pages}{41} (\bibinfo{year}{2002}).

\bibitem[{\citenamefont{Vavro}(2001)}]{T2001}
\bibinfo{author}{\bibfnamefont{J.}~\bibnamefont{Vavro}},
  \bibinfo{journal}{Phys. Rev. E} \textbf{\bibinfo{volume}{63}},
  \bibinfo{pages}{057104} (\bibinfo{year}{2001}).

\bibitem[{\citenamefont{Zimmerman and Pan}(2000)}]{E2000}
\bibinfo{author}{\bibfnamefont{F.~M.} \bibnamefont{Zimmerman}}
  \bibnamefont{and} \bibinfo{author}{\bibfnamefont{X.}~\bibnamefont{Pan}},
  \bibinfo{journal}{Phys. Rev. Lett.} \textbf{\bibinfo{volume}{85}},
  \bibinfo{pages}{618} (\bibinfo{year}{2000}).

\bibitem[{\citenamefont{Yilmaz et~al.}(2004)\citenamefont{Yilmaz, Rajagopal,
  and Zimmerman}}]{E2004}
\bibinfo{author}{\bibfnamefont{M.~B.} \bibnamefont{Yilmaz}},
  \bibinfo{author}{\bibfnamefont{A.}~\bibnamefont{Rajagopal}},
  \bibnamefont{and} \bibinfo{author}{\bibfnamefont{F.~M.}
  \bibnamefont{Zimmerman}}, \bibinfo{journal}{Phys. Rev. B}
  \textbf{\bibinfo{volume}{69}}, \bibinfo{pages}{125413}
  (\bibinfo{year}{2004}).

\bibitem[{\citenamefont{Maniwa et~al.}(2003)\citenamefont{Maniwa, Kataura,
  Matsuda, and Okabe}}]{E2003}
\bibinfo{author}{\bibfnamefont{Y.}~\bibnamefont{Maniwa}},
  \bibinfo{author}{\bibfnamefont{H.}~\bibnamefont{Kataura}},
  \bibinfo{author}{\bibfnamefont{K.}~\bibnamefont{Matsuda}}, \bibnamefont{and}
  \bibinfo{author}{\bibfnamefont{Y.}~\bibnamefont{Okabe}},
  \bibinfo{journal}{New J. Phys.} \textbf{\bibinfo{volume}{5}},
  \bibinfo{pages}{127} (\bibinfo{year}{2003}).

\bibitem[{\citenamefont{Wartell and Benight}(1985)}]{E1985}
\bibinfo{author}{\bibfnamefont{R.~M.} \bibnamefont{Wartell}} \bibnamefont{and}
  \bibinfo{author}{\bibfnamefont{A.~S.} \bibnamefont{Benight}},
  \bibinfo{journal}{Phys. Rep.} \textbf{\bibinfo{volume}{126}},
  \bibinfo{pages}{67} (\bibinfo{year}{1985}).

\end{thebibliography}

\end{document}